\documentclass[aps,prb,showpacs,twocolumn,groupedaddress]{revtex4}
\usepackage{graphicx}

\begin{document}
\bibliographystyle{apsrev}

\title{Full optical responses of one-dimensional metallic photonic crystal 
slabs}

\author{Masanobu Iwanaga}
\email[Electronic address: ]{iwanaga@phys.tohoku.ac.jp}
\affiliation{Department of Physics, Graduate School of Science, 
Tohoku University, Sendai 980-8578, Japan}
\author{Teruya Ishihara}
\affiliation{Department of Physics, Graduate School of Science, 
Tohoku University, Sendai 980-8578, Japan}
\affiliation{Institute for Physical and Chemical Research (RIKEN), 
Wako 351-0198, Japan}

\date{\today}

\begin{abstract}
We reveal all the linear optical responses, reflection, transmission, 
and, diffraction, of typical one-dimensional 
metallic photonic crystal slabs (MPhCS) with the periodicity of a half 
micrometer. Maxwell equations for the structure of deep grooves are 
solved numerically with good precision by using the formalism of 
scattering matrix, without assuming perfect conductivity. We verify 
characteristic optical properties such as nearly perfect transmission 
and reflection. Moreover, we present large reflective diffraction and 
show that, in the energy range where diffraction channels 
are open, the photonic states in the MPhCS 
originate from surface plasmon polaritons. 
\end{abstract}


\maketitle

Electromagnetic (EM) waves in artificial structures were one of the old 
problems\cite{Rayleigh,Bethe} and had been thought to be already resolved. 
Since the observation of extraordinary transmission of perforated 
silver film in 1998,\cite{Ebbesen} 
optical responses of metallic photonic crystal slabs 
(MPhCS) have attracted much interest as a novel phenomenon 
involving surface plasmons. 
Many researchers have been stimulated by the phenomenon 
and have tried to explain the properties of transmission in MPhCS. 

\begin{figure}
\includegraphics[width=6.5cm]{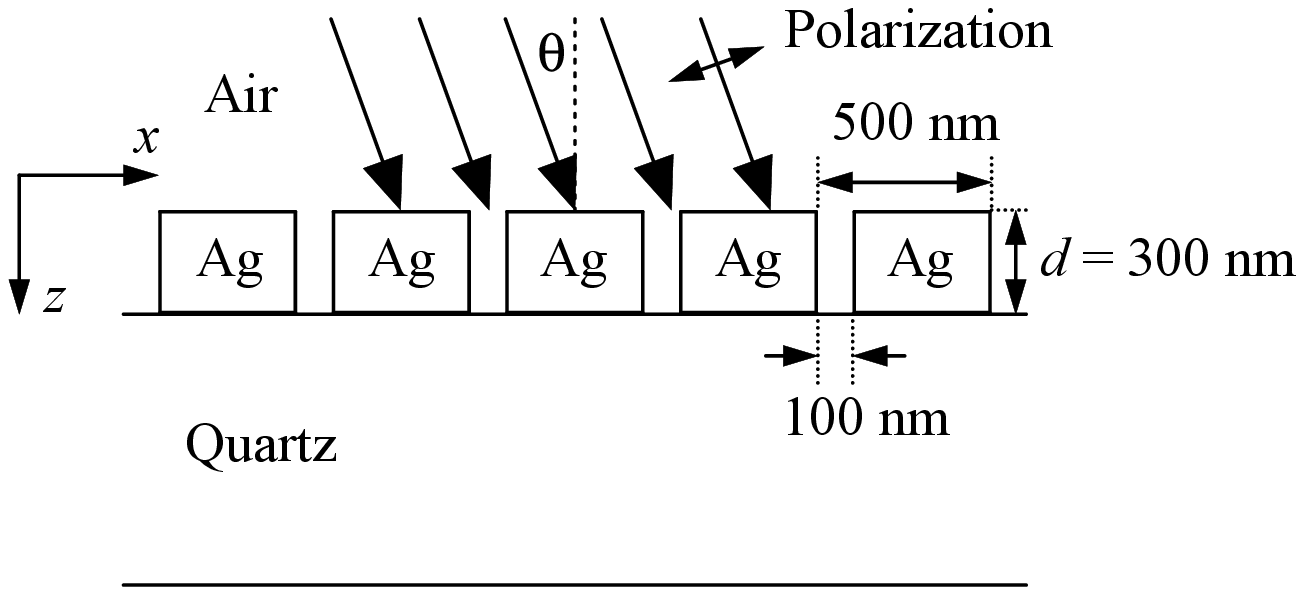}%
\caption{Structure of 1D MPhCS; the periodicity is 500 nm, 
the width of air slit is 100 nm, and the thickness $d$ is 300 nm. 
The MPhCS is on the substrate of quartz. Incident light (arrows) 
travels in the $xz$ plane. Incident angle $\theta$ 
is the angle between the normal and incident directions. The 
incident light is $p$-polarized.}
\label{fig1}
\end{figure}

Even when we restrict an object to a one-dimensional (1D) MPhCS 
as drawn in Fig.\ \ref{fig1}, 
 the zeroth-order transmission was 
already discussed in not a few reports\cite{Porto,Popov2,Vidal} 
and explained qualitatively. 
In order to evaluate the transmission numerically, 
Maxwell equations were solved with additional assumptions such as 
perfect conductivity.\cite{Porto,Vidal} 
In those reports, transmission coefficient was focused on and calculated. 

On the other hand, diffraction has not been fully analyzed, 
to our best knowledge, although the surface plasmons in MPhCS are 
usually connected to diffraction. In fact, diffraction in optically 
thick metallic grooves such as Fig.\ \ref{fig1} 
has been a tough subject since the first era of surface plasmons 
in 1980s. Maxwell equations for metallic grooves were analytically solved 
under the assumption of perfect conductivity;\cite{Sheng} 
however, the results based on the hypothesis deviate from 
experimental results quantitatively. 
Thus, there has been strong motivation for describing the optical responses 
of MPhCS quantitatively. 

The formalism to calculate full optical response including diffraction 
was tried to construct, independently of the extensive studies on 
extraordinary transmission.\cite{Moharam,Lalanne,Popov1,Ager,Li3,Whittaker} 
To include diffractive components, scattering-matrix 
(S-Matrix) formalism is suitable.\cite{Popov1,Ager,Li3,Whittaker,Tikh} 
The S-Matrix method is a kind of transfer-matrix method for 
the propagation of EM wave. The method was invented to 
ensure numerical stability.\cite{Ko} 
We adopt the S-Matrix formalism; 
it is based on the ordinary local response of EM fields, and 
the explicit expression for optical responses 
was presented in Ref.\ \onlinecite{Tikh}. The formalism worked quite well 
for thin 1D gold PhCS.\cite{Christ2} 

In this Communication, we solve Maxwell equations for the 1D MPhCS 
accurately and evaluate all the linear optical responses. 
In particular, we explore the large 
diffraction and seek the physical insight. 
Furthermore, we discuss the physical meaning of this computation. 

Figure \ref{fig1} shows the structure we mainly examine here. 
Rectangular silver rods are infinitely long and 
periodically located at the 100-nm interval 
on the quartz substrate. Incident light is $p$-polarized, that is, 
the polarization of electric field is perpendicular to the axes of Ag rods. 
Setting $x$ axis to be the periodic direction, 
the electric field of $p$-polarized incident light 
is expanded as 
\begin{eqnarray}
\textbf{E}(\textbf{r},t) &=& \sum_m \textbf{E}_{m} \exp%
(i \textbf{k}_{m}\cdot\textbf{r} - i \omega t), 
\label{E_expand} \\ 
\textbf{k}_{m} &=&(k_{x,m}, k_y, k_z), \nonumber\\
k_{x,m} &=& k_x + 2\pi m/a, 
(m = 0, \pm 1, \pm 2, \cdots), \nonumber\\
k_x &=& k_0\sin\theta, \nonumber\\
k_z &=& \sqrt{k_0^2 - k_{x,m}^2 -k_y^2} \nonumber
\end{eqnarray}
where $a$ is the periodicity, 
$k_0$ is wavenumber of incident light, $\theta$ is defined in 
Fig.\ \ref{fig1}, and $k_y = 0$. 
The electric fields in the MPhCS and substrate 
are expressed accordingly. The magnetic fields are described 
likewise. In the present S-Matrix formalism, 
the infinite expansion of Eq.\ (\ref{E_expand}) gives the Fourier-based 
linear equations which are exactly equivalent to Maxwell equations. 
Practically, the truncation of expansion is inevitable. 
We cut Eq.\ (\ref{E_expand}) at $m = \pm m_{\rm max}$ and then 
the total number of harmonics $N_{\rm total}$ is $2m_{\rm max}+1$. 
Numerical computations were executed in the highly vectorized 
supercomputers, which enable to make 32 parallelization. 
The complex dielectric constant of silver is taken from Ref.\ 
\onlinecite{Johnson}. 

\begin{figure}
\includegraphics[width=6.5cm]{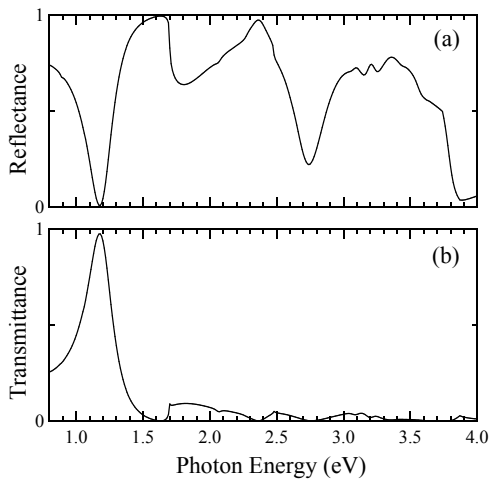}%
\caption{(a) Reflectance and (b) transmittance spectra of the 1D MPhCS 
at incident angle $\theta = 0^{\circ}$. 
}
\label{fig2}
\end{figure}

Figure \ref{fig2} displays (a) reflectance and (b) transmittance spectra 
calculated at incident angle $\theta = 0^{\circ}$ and 
$N_{\rm total} = 1265$. Diffraction is calculated simultaneously in this 
method. The fluctuation of the calculated values is within 1\% 
for $N_{\rm total} > 1200$ in this energy range. 

Nearly perfect transmission at 1.2 eV in Fig.\ \ref{fig2}(b) is due to 
the waveguide mode in the air slit. 
The qualitative analysis of the mode was reported previously%
\cite{Porto,Popov2,Vidal} and the Fabry-P\'erot nature was pointed 
out. The property is reproduced in our calculation. 
The waveguide mode does not appear above 1.7 eV where diffraction is 
effective, and transmission is not strongly enhanced. 
Reflectance is nearly 100\% at 1.63 and 2.36 eV in Fig.\ \ref{fig2}(a); 
the response is called Wood anomaly.\cite{Wood} It is usually observed 
at the resonances of surface plasmons in metallic grooves. 

We note here the reason why the large $N_{\rm total}$ is required. 
This is mainly because there exist the resonances under $p$ polarization, 
due to the waveguide mode and surface plasmons. 
In a good metal such as silver, the resonances induce 
the spatially modulated and intense local EM fields in and near the MPhCS; 
the situation is in general hard to express with a small $N_{\rm total}$. 
Therefore, reliable numerical 
calculation for optical resonances requires the large $N_{\rm total}$ and 
actually becomes quite demanding. However, we can achieve optical 
responses with good precision by using a large $N_{\rm total}$ 
in a wide frequency range from near infrared to ultraviolet. 
On the contrary, in the off-resonant frequency range, 
far less $N_{\rm total}$ is enough to obtain the solution of Maxwell 
equations. 
Indeed, we can calculate the optical responses within 0.2\% fluctuation 
by using a $N_{\rm total}$ of a few tens under $s$ polarization, 
when the polarization of electric field is parallel to the rod axes. 

\begin{figure}
\includegraphics[width=6.5cm]{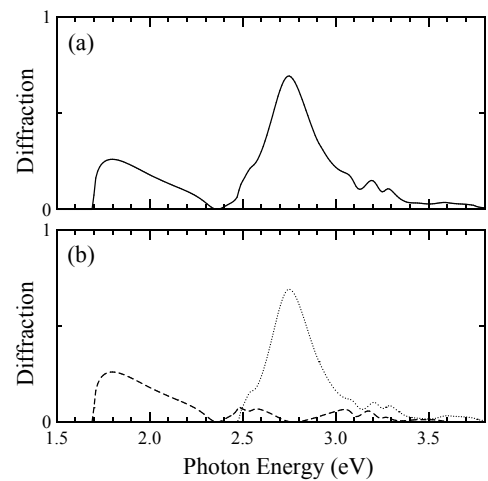}%
\caption{(a) Solid line: total diffraction spectrum of the 1D MPhCS. 
(b) Dashed line: transmissive diffraction. 
Dotted line: reflective diffraction. 
The sum of these diffractions is the total diffraction. 
The incident angle $\theta$ is $0^{\circ}$. 
}
\label{fig3}
\end{figure}

Figure \ref{fig3}(a) presents total diffraction spectrum under the 
normal incidence (solid line). 
We use the conventional notation in diffraction theory; 
$R_n$ means the $n$th reflective 
diffraction and $T_n$ the $n$th transmissive diffraction. 
Then, reflectance is $R_0$, transmittance is $T_0$, and 
the total diffraction is $\sum_{n \neq 0} (R_n + T_n)$. 
Dashed line in Fig.\ \ref{fig3}(b) shows transmissive diffraction 
$\sum_{n \neq 0} T_{n}$. Similarly, 
the reflective diffraction $\sum_{n \neq 0} R_{n}$ 
is represented with dotted line. 

The components of diffraction have the following properties. 
The transmissive diffraction $T_{\pm 1}$ emerges at 1.70 eV. 
The sharp onset corresponds to the opening of 
the first-order channel of Bragg diffraction at the interface between 
the quartz and MPhCS. The second channel opens at 3.40 eV and induces 
$T_{\pm 2}$. On the other hand, the reflective diffraction appears 
above 2.48 eV and is composed of $R_{\pm 1}$ in Fig.\ \ref{fig3}(b). 
The onset is due to the opening of the first-order channel of Bragg 
diffraction at the interface between the MPhCS and air. 
The sum of $R_{\pm 1}$ reaches the maximum of 70\% at 2.75 eV. 
The spectra of diffraction are discussed later. 

\begin{figure}
\includegraphics[width=6.5cm]{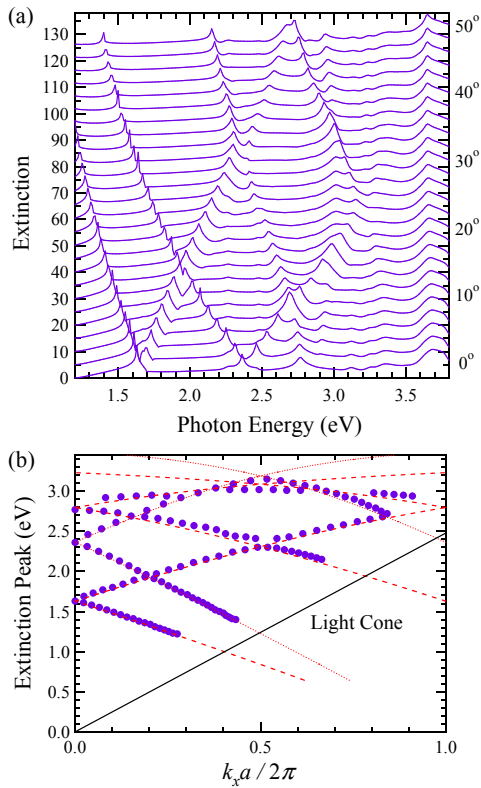}%
\caption{(a) Extinction spectra of the 1D MPhCS at incident angles 
from 0 to 50 degrees. 
(b) Dispersion relation of extinction peaks (solid circles). 
Dotted and dashed lines: the dispersions of reduced SPPs at the air-MPhCS 
and the quartz-MPhCS interfaces, respectively. 
The periodicity $a$ is 500 nm. 
The wavenumber $k_x$ is described in the text 
after Eq.\ (\ref{dispersion}). 
}
\label{fig4}
\end{figure}

To examine the dispersion of resonances above 1.3 eV, 
we show extinction spectra 
at incident angle $\theta=0$ to 50 degrees in Fig.\ \ref{fig4}(a). 
Extinction is $-\ln(T_0)$. 
The peaks of the extinction spectra are the minima of transmittance by 
definition and plotted with solid circles in Fig.\ \ref{fig4}(b). 

We compare the dispersion of extinction with that of surface plasmon 
polariton (SPP). SPP at flat interface is generally expressed by 
\begin{equation}
\omega = ck_{\parallel} \sqrt{\frac{1}{\varepsilon} + %
\frac{1}{\varepsilon_{\rm Ag}(\omega)}}
\label{dispersion}
\end{equation}
where $\omega$ is the frequency of incident light, 
$c$ is the velocity of light in vacuum, and $k_{\parallel}$ 
is wavenumber of SPP along the $x$ axis, 
$\varepsilon$ is dielectric constant of air or quartz, 
and $\varepsilon_{\rm Ag}$ is the complex dielectric 
constant of silver. Note that when 
$\varepsilon_{\rm Ag}$ is complex number, $k_{\parallel}$ 
is also complex. For silver in the energy range of interest, 
Eq.\ (\ref{dispersion}) holds with replacing $\varepsilon_{\rm Ag}$ by 
Re($\varepsilon_{\rm Ag}$) and $k_{\parallel}$ by Re($k_{\parallel}$).%
\cite{Raether} We set $k_x = \mathrm{Re}(k_{\parallel})$ 
in accordance with Eq.\ (\ref{E_expand}).
The dispersion of SPP is evaluated and shown in 
Fig.\ \ref{fig4}(b) after reducing into the Brillouin zone. 
The dotted and dashed lines represent the SPPs at the air-Ag and 
the quartz-Ag interfaces, respectively. 

The dispersion of extinction is in good agreement with the dispersion 
of reduced SPPs. The agreement indicates that the reduced SPPs are 
resonantly excited, and strongly suggests that the photonic states 
in the MPhCS have formed from the reduced SPPs above 1.3 eV. 
The agreement is also consistent with the previous analysis with 
the method of rigorously coupled wave.\cite{Cao} 

As shown in Fig.\ \ref{fig3}, 
the sharp onset of diffraction at 1.70 eV corresponds to the opening of 
$T_{\pm 1}$. 
On the other hand, as shown in Figs.\ \ref{fig3} and \ref{fig4}, 
the sum of $R_{\pm 1}$ gradually increases and has the maximum 
at 2.75 eV; the maximum is close to the resonance of the second-order SPP 
at 2.80 eV and $k_x=0$. How can we explain the enhancement of diffraction? 
To extract more information, 
we examine the diffraction with varying only the thickness of MPhCS. 

\begin{figure}
\includegraphics[width=7.0cm]{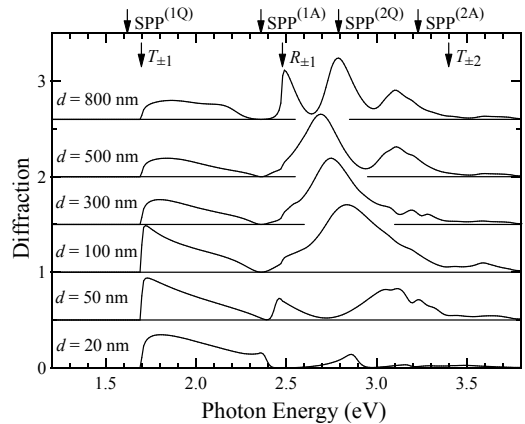}%
\caption{Total diffraction spectra at various thickness $d$. 
The periodicity and the width of air slit are shown in Fig.\ \ref{fig1}. 
The incident angle $\theta$ is $0^{\circ}$. 
Arrows indicate the opening of $n$th diffraction ($n=\pm 1, \pm 2$) 
and the resonances of SPPs. For example, 
SPP$^{({\rm 1Q})}$ and SPP$^{({\rm 1A})}$ stand for the first-order 
resonances at the quartz-MPhCS and air-MPhCS interfaces, respectively. 
The energies are taken from Fig.\ \ref{fig4}(b). 
The spectra at $d\ge50$ nm are offset for clarity.}
\label{fig5}
\end{figure}

Figure \ref{fig5} represents total diffraction spectra at the thickness 
$d=20$, 50, 100, 300, 500, and 800 nm under the normal incidence. 
Arrows indicate the energy positions at which the channels of diffraction 
open, and the resonances of reduced SPPs. 
SPP$^{(n{\rm Q})}$ and SPP$^{(n{\rm A})}$ denote the $n$th resonances 
at the quartz-MPhCS and air-MPhCS interfaces, respectively. 
The positions of SPP resonances are taken from the dispersion in 
Fig.\ \ref{fig4}(b). 
In all the spectra, diffraction arises at 1.7 eV and is comprised of 
$T_{\pm 1}$. At 1.7--2.4 eV, diffraction reaches a few tens of percents 
and the spectra show similar tendency irrespective of the thickness. 
Above 2.4 eV, the diffraction spectra depend on the thickness. 
In the thin MPhCS of $d = 20$ and 50 nm, the components $T_{\pm 1}$ and 
$R_{\pm 1}$ coexist, and are not prominently enhanced at 2.7--2.8 eV. 
In the optically thick MPhCS of $d \ge 100$ nm, large peaks of diffraction 
appear at 2.7--2.8 eV and are dominantly composed of $R_{\pm 1}$; 
see Fig.\ \ref{fig3}(b) for the case of $d = 300$ nm. 
The peak positions change with the thickness, while the extinction 
spectra show that the energies of SPP resonances are common in the MPhCS of 
$d \ge 100$ nm. Thus, though the large peaks at 2.7--2.8 eV are close to 
the resonance SPP$^{\rm (2Q)}$ at 2.80 eV, they cannot be simply ascribed 
to the resonant enhancement by reduced SPPs. 

By varying the thickness from $d=20$ to 800 nm, one can observe that 
the diffraction spectra change in a quite complicated way. 
The changes in the diffraction spectra probably come from 
the changes of the photonic states in the MPhCS. 
We believe that the prominent peaks of diffraction indicate 
the resonant emission modes in the MPhCS. 

To find the resonant modes in the MPhCS, 
it is a key to solve the equation of 
det$(\mathcal{S})=0$, where $\mathcal{S}$ is 
a S-Matrix for the 1D MPhCS, satisfying 
$|in\rangle = \mathcal{S}|out\rangle$, and $|in\rangle$ 
and $|out\rangle$ are input and output states, respectively. 
It is surely demanding and a future issue to solve det$(\mathcal{S})=0$ 
for a large $N_{\rm total}$. The solutions of the equation and 
the photonic modes were obtained for the two-dimensional PhCS made of 
dielectrics with a small $N_{\rm total}$.\cite{Tikh,Tikh2} 

We have obtained a numerically accurate solution of Maxwell equations 
for the 1D MPhCS and have shown the optical responses. We
would like to compare the present results with measured ones 
quantitatively. For the precise comparison of the present 
calculated results with measurement, 
it is better to compute with using the complex dielectric 
constants obtained from the material used in the measurement, 
because the values can change under the condition of 
fabrication and environment.\cite{Kovacs} 
Such comparison is a test for the validity of 
the ordinary local response of EM fields such as the electric flux density 
$\mathbf{D}(\mathbf{r}) = \varepsilon(\mathbf{r})\mathbf{E}(\mathbf{r})$. 
If there exists definite discrepancy beyond the errors in the numerical 
calculation and experiment, it would be necessary to take into account 
the microscopic nonlocal response of EM fields.\cite{Cho} 
It is an interesting subject to find the metallic nano-structures 
in which the nonlocal responses emerge distinctively. 

In the S-Matrix formalism, one can obtain absorption $A$ by the 
relation of $A = 1 - \sum_{{\rm all}\,n}(R_n + T_n)$. 
Thus, the absorption can be evaluated from all the 
components in optical measurement. 
The absorption spectrum is also measured directly 
by photoacoustic method;\cite{Inagaki,Iwanaga} 
the method is useful for diffractive media. 

At the end of discussion we mention the previous reports. 
Nearly perfect transmission, which appears at 1.2 eV in Fig.\ 
\ref{fig2}(b), was attributed to the waveguide mode in the slit 
mainly from the distribution of EM fields.\cite{Popov2} 
The high-efficient transmission was also analyzed under the assumption of 
the perfect conductivity only in the slit.\cite{Porto,Vidal} 
Besides, the coupling of SPP with light was recently analyzed with 
extracting a effective term of S-Matrix and the transmission was explained 
qualitatively.\cite{Lee} 
In this Communication, Maxwell equations for the 1D MPhCS have been solved 
numerically without any further technical assumption except for 
the inevitable truncation. From the solution, 
the optical responses have been evaluated directly in a wide frequency range 
from infrared to ultraviolet. 

In conclusion, we selected typical 1D MPhCS made of silver and 
have computed the full optical responses precisely using the measured 
complex dielectric constant in the S-Matrix formalism. 
This is the first achievement concerning realistic calculation for 1D MPhCS 
with the structure of deep groove. 
We have confirmed the resonant excitation of reduced SPPs and found that 
the large reflective diffraction appears in the optically thick MPhCS. 
The origin has been discussed in view of resonant enhancement by reduced 
SPPs and of resonant emission modes in the MPhCS. 
The present computation, assuming the local response 
of EM fields, can be readily compared with measurement. 

\begin{acknowledgments}
We would like to acknowledge S.\ G.\ Tikhodeev for helpful discussion. 
The numerical implementation was partially supported by Information 
Synergy Center, Tohoku University. This study was supported in part 
by a Grant-in-Aid for Scientific Research by 
the Ministry of Education, Culture, Sports, Science, and 
Technology, Japan.
\end{acknowledgments}

\bibliography{iwanaga7}

\begin{thebibliography}{26}
\expandafter\ifx\csname natexlab\endcsname\relax\def\natexlab#1{#1}\fi
\expandafter\ifx\csname bibnamefont\endcsname\relax
  \def\bibnamefont#1{#1}\fi
\expandafter\ifx\csname bibfnamefont\endcsname\relax
  \def\bibfnamefont#1{#1}\fi
\expandafter\ifx\csname citenamefont\endcsname\relax
  \def\citenamefont#1{#1}\fi
\expandafter\ifx\csname url\endcsname\relax
  \def\url#1{\texttt{#1}}\fi
\expandafter\ifx\csname urlprefix\endcsname\relax\def\urlprefix{URL }\fi
\providecommand{\bibinfo}[2]{#2}
\providecommand{\eprint}[2][]{\url{#2}}

\bibitem[{\citenamefont{{Lord Rayleigh}}(1907)}]{Rayleigh}
\bibinfo{author}{\bibnamefont{{Lord Rayleigh}}}, \bibinfo{journal}{{Proc.\ R.\
  Soc.\ London Ser.\ A }} \textbf{\bibinfo{volume}{79}}, \bibinfo{pages}{399}
  (\bibinfo{year}{1907}).

\bibitem[{\citenamefont{{H.\ A.\ Bethe}}(1944)}]{Bethe}
\bibinfo{author}{\bibnamefont{{H.\ A.\ Bethe}}}, \bibinfo{journal}{{Phys.\
  Rev.\ }} \textbf{\bibinfo{volume}{66}}, \bibinfo{pages}{163}
  (\bibinfo{year}{1944}).

\bibitem[{\citenamefont{{T.\ W.\ Ebbesen} et~al.}(1998)\citenamefont{{T.\ W.\
  Ebbesen}, {H.\ J.\ Lezec}, {H.\ F.\ Ghaemi}, {T.\ Thio}, and {P.\ A.\
  Wolff}}}]{Ebbesen}
\bibinfo{author}{\bibnamefont{{T.\ W.\ Ebbesen}}},
  \bibinfo{author}{\bibnamefont{{H.\ J.\ Lezec}}},
  \bibinfo{author}{\bibnamefont{{H.\ F.\ Ghaemi}}},
  \bibinfo{author}{\bibnamefont{{T.\ Thio}}}, \bibnamefont{and}
  \bibinfo{author}{\bibnamefont{{P.\ A.\ Wolff}}}, \bibinfo{journal}{{Nature }}
  \textbf{\bibinfo{volume}{391}}, \bibinfo{pages}{667} (\bibinfo{year}{1998}).

\bibitem[{\citenamefont{{J.\ A.\ Porto} et~al.}(1999)\citenamefont{{J.\ A.\
  Porto}, {F.\ J.\ Garc\'{\i}a-Vidal}, and {J.\ B.\ Pendry}}}]{Porto}
\bibinfo{author}{\bibnamefont{{J.\ A.\ Porto}}},
  \bibinfo{author}{\bibnamefont{{F.\ J.\ Garc\'{\i}a-Vidal}}},
  \bibnamefont{and} \bibinfo{author}{\bibnamefont{{J.\ B.\ Pendry}}},
  \bibinfo{journal}{{Phys.\ Rev.\ Lett.\ }} \textbf{\bibinfo{volume}{83}},
  \bibinfo{pages}{2845} (\bibinfo{year}{1999}).

\bibitem[{\citenamefont{{E.\ Popov} et~al.}(2000)\citenamefont{{E.\ Popov},
  {M.\ Nevi\`{e}re}, and {R.\ Reinisch}}}]{Popov2}
\bibinfo{author}{\bibnamefont{{E.\ Popov}}}, \bibinfo{author}{\bibnamefont{{M.\
  Nevi\`{e}re}}}, \bibnamefont{and} \bibinfo{author}{\bibnamefont{{R.\
  Reinisch}}}, \bibinfo{journal}{{Phys.\ Rev.\ B }}
  \textbf{\bibinfo{volume}{62}}, \bibinfo{pages}{16100} (\bibinfo{year}{2000}),
  \bibinfo{note}{and earlier references cited therein}.

\bibitem[{\citenamefont{{F.\ J.\ Garc\'{\i}a-Vidal} and {L.\
  Mart\'{\i}n-Moreno}}(2002)}]{Vidal}
\bibinfo{author}{\bibnamefont{{F.\ J.\ Garc\'{\i}a-Vidal}}} \bibnamefont{and}
  \bibinfo{author}{\bibnamefont{{L.\ Mart\'{\i}n-Moreno}}},
  \bibinfo{journal}{{Phys.\ Rev.\ B }} \textbf{\bibinfo{volume}{66}},
  \bibinfo{pages}{155412} (\bibinfo{year}{2002}).

\bibitem[{\citenamefont{{P.\ Sheng} et~al.}(1982)\citenamefont{{P.\ Sheng},
  {R.\ S.\ Stepleman}, and {P.\ N.\ Sanda}}}]{Sheng}
\bibinfo{author}{\bibnamefont{{P.\ Sheng}}}, \bibinfo{author}{\bibnamefont{{R.\
  S.\ Stepleman}}}, \bibnamefont{and} \bibinfo{author}{\bibnamefont{{P.\ N.\
  Sanda}}}, \bibinfo{journal}{{Phys.\ Rev.\ B }} \textbf{\bibinfo{volume}{26}},
  \bibinfo{pages}{2907} (\bibinfo{year}{1982}), \bibinfo{note}{and ealier
  references cited therein}.

\bibitem[{\citenamefont{{E.\ Popov} et~al.}(1986)\citenamefont{{E.\ Popov},
  {L.\ Mashev}, and {D.\ Maystre}}}]{Popov1}
\bibinfo{author}{\bibnamefont{{E.\ Popov}}}, \bibinfo{author}{\bibnamefont{{L.\
  Mashev}}}, \bibnamefont{and} \bibinfo{author}{\bibnamefont{{D.\ Maystre}}},
  \bibinfo{journal}{{Opt.\ Acta }} \textbf{\bibinfo{volume}{33}},
  \bibinfo{pages}{607} (\bibinfo{year}{1986}).

\bibitem[{\citenamefont{{C.\ D.\ Ager} and {H.\ P.\ Hughes}}(1991)}]{Ager}
\bibinfo{author}{\bibnamefont{{C.\ D.\ Ager}}} \bibnamefont{and}
  \bibinfo{author}{\bibnamefont{{H.\ P.\ Hughes}}}, \bibinfo{journal}{{Phys.\
  Rev.\ B }} \textbf{\bibinfo{volume}{44}}, \bibinfo{pages}{13452}
  (\bibinfo{year}{1991}).

\bibitem[{\citenamefont{{L.\ Li}}(1996)}]{Li3}
\bibinfo{author}{\bibnamefont{{L.\ Li}}}, \bibinfo{journal}{{J.\ Opt.\ Soc.\
  Am.\ A }} \textbf{\bibinfo{volume}{13}}, \bibinfo{pages}{1024}
  (\bibinfo{year}{1996}).

\bibitem[{\citenamefont{{D.\ M.\ Whittaker} and {I.\ S.\
  Culshaw}}(1999)}]{Whittaker}
\bibinfo{author}{\bibnamefont{{D.\ M.\ Whittaker}}} \bibnamefont{and}
  \bibinfo{author}{\bibnamefont{{I.\ S.\ Culshaw}}}, \bibinfo{journal}{{Phys.\
  Rev.\ B }} \textbf{\bibinfo{volume}{60}}, \bibinfo{pages}{2610}
  (\bibinfo{year}{1999}).

\bibitem[{\citenamefont{{M.\ G.\ Moharam} et~al.}(1995)\citenamefont{{M.\ G.\
  Moharam}, {E.\ B.\ Grann}, {D.\ A.\ Pommet}, and {T.\ K.\
  Gaylord}}}]{Moharam}
\bibinfo{author}{\bibnamefont{{M.\ G.\ Moharam}}},
  \bibinfo{author}{\bibnamefont{{E.\ B.\ Grann}}},
  \bibinfo{author}{\bibnamefont{{D.\ A.\ Pommet}}}, \bibnamefont{and}
  \bibinfo{author}{\bibnamefont{{T.\ K.\ Gaylord}}}, \bibinfo{journal}{{J.\
  Opt.\ Soc.\ Am.\ A }} \textbf{\bibinfo{volume}{12}}, \bibinfo{pages}{1068}
  (\bibinfo{year}{1995}).

\bibitem[{\citenamefont{{P.\ Lalanne} and {G.\ M.\ Morris}}(1996)}]{Lalanne}
\bibinfo{author}{\bibnamefont{{P.\ Lalanne}}} \bibnamefont{and}
  \bibinfo{author}{\bibnamefont{{G.\ M.\ Morris}}}, \bibinfo{journal}{{J.\
  Opt.\ Soc.\ Am.\ A }} \textbf{\bibinfo{volume}{13}}, \bibinfo{pages}{779}
  (\bibinfo{year}{1996}).

\bibitem[{\citenamefont{{S.\ G.\ Tikhodeev} et~al.}(2002)\citenamefont{{S.\ G.\
  Tikhodeev}, {A.\ L.\ Yablonskii}, {E.\ A.\ Muljarov}, {N.\ A.\ Gippius}, and
  {T.\ Ishihara}}}]{Tikh}
\bibinfo{author}{\bibnamefont{{S.\ G.\ Tikhodeev}}},
  \bibinfo{author}{\bibnamefont{{A.\ L.\ Yablonskii}}},
  \bibinfo{author}{\bibnamefont{{E.\ A.\ Muljarov}}},
  \bibinfo{author}{\bibnamefont{{N.\ A.\ Gippius}}}, \bibnamefont{and}
  \bibinfo{author}{\bibnamefont{{T.\ Ishihara}}}, \bibinfo{journal}{{Phys.\
  Rev.\ B }} \textbf{\bibinfo{volume}{66}}, \bibinfo{pages}{045102}
  (\bibinfo{year}{2002}).

\bibitem[{\citenamefont{{D.\ Y.\ K.\ Ko} and {J.\ C.\ Inkson}}(1988)}]{Ko}
\bibinfo{author}{\bibnamefont{{D.\ Y.\ K.\ Ko}}} \bibnamefont{and}
  \bibinfo{author}{\bibnamefont{{J.\ C.\ Inkson}}}, \bibinfo{journal}{{Phys.\
  Rev.\ B }} \textbf{\bibinfo{volume}{38}}, \bibinfo{pages}{9945}
  (\bibinfo{year}{1988}).

\bibitem[{\citenamefont{{A.\ Christ} et~al.}(2004)\citenamefont{{A.\ Christ},
  {T.\ Zentgraf}, {J.\ Kuhl}, {S.\ G.\ Tikhodeev}, {N.\ A.\ Gippius}, and {H.\
  Giessen}}}]{Christ2}
\bibinfo{author}{\bibnamefont{{A.\ Christ}}},
  \bibinfo{author}{\bibnamefont{{T.\ Zentgraf}}},
  \bibinfo{author}{\bibnamefont{{J.\ Kuhl}}},
  \bibinfo{author}{\bibnamefont{{S.\ G.\ Tikhodeev}}},
  \bibinfo{author}{\bibnamefont{{N.\ A.\ Gippius}}}, \bibnamefont{and}
  \bibinfo{author}{\bibnamefont{{H.\ Giessen}}}, \bibinfo{journal}{{Phys.\
  Rev.\ B }} \textbf{\bibinfo{volume}{70}}, \bibinfo{pages}{125113}
  (\bibinfo{year}{2004}).

\bibitem[{\citenamefont{{P.\ B.\ Johonson} and {R.\ W.\
  Christy}}(1972)}]{Johnson}
\bibinfo{author}{\bibnamefont{{P.\ B.\ Johonson}}} \bibnamefont{and}
  \bibinfo{author}{\bibnamefont{{R.\ W.\ Christy}}}, \bibinfo{journal}{{Phys.\
  Rev.\ B }} \textbf{\bibinfo{volume}{6}}, \bibinfo{pages}{4370}
  (\bibinfo{year}{1972}).

\bibitem[{\citenamefont{{R.\ W.\ Wood}}(1902)}]{Wood}
\bibinfo{author}{\bibnamefont{{R.\ W.\ Wood}}}, \bibinfo{journal}{{Philos.\
  Mag.\ }} \textbf{\bibinfo{volume}{4}}, \bibinfo{pages}{396}
  (\bibinfo{year}{1902}).

\bibitem[{\citenamefont{{H.\ Raether}}(1988)}]{Raether}
\bibinfo{author}{\bibnamefont{{H.\ Raether}}}, \emph{\bibinfo{title}{Surface
  Plasmons on Smooth and Rough Surfaces and on Gratings}}
  (\bibinfo{publisher}{Springer}, \bibinfo{address}{Berlin},
  \bibinfo{year}{1988}).

\bibitem[{\citenamefont{{Q.\ Cao} and {P.\ Lalanne}}(2002)}]{Cao}
\bibinfo{author}{\bibnamefont{{Q.\ Cao}}} \bibnamefont{and}
  \bibinfo{author}{\bibnamefont{{P.\ Lalanne}}}, \bibinfo{journal}{{Phys.\
  Rev.\ Lett.\ }} \textbf{\bibinfo{volume}{88}}, \bibinfo{pages}{057403}
  (\bibinfo{year}{2002}).

\bibitem[{\citenamefont{{S.\ G.\ Tikhodeev} et~al.}(2005)\citenamefont{{S.\ G.\
  Tikhodeev}, {N.\ A.\ Gippus}, {A.\ Christ}, {T.\ Zentgraf}, {J.\ Kuhl}, and
  {H.\ Giessen}}}]{Tikh2}
\bibinfo{author}{\bibnamefont{{S.\ G.\ Tikhodeev}}},
  \bibinfo{author}{\bibnamefont{{N.\ A.\ Gippus}}},
  \bibinfo{author}{\bibnamefont{{A.\ Christ}}},
  \bibinfo{author}{\bibnamefont{{T.\ Zentgraf}}},
  \bibinfo{author}{\bibnamefont{{J.\ Kuhl}}}, \bibnamefont{and}
  \bibinfo{author}{\bibnamefont{{H.\ Giessen}}}, \bibinfo{journal}{{Phys.\
  Status Solidi C }} \textbf{\bibinfo{volume}{2}}, \bibinfo{pages}{795}
  (\bibinfo{year}{2005}).

\bibitem[{\citenamefont{{G.\ J.\ Kovacs}}(1978)}]{Kovacs}
\bibinfo{author}{\bibnamefont{{G.\ J.\ Kovacs}}}, \bibinfo{journal}{{Surf.\
  Sci.\ }} \textbf{\bibinfo{volume}{78}}, \bibinfo{pages}{L245}
  (\bibinfo{year}{1978}).

\bibitem[{\citenamefont{{K.\ Cho}}(2003)}]{Cho}
\bibinfo{author}{\bibnamefont{{K.\ Cho}}}, \emph{\bibinfo{title}{Optical
  Response of Nanostructures: Microscopic Nonlocal Theory}}
  (\bibinfo{publisher}{Springer}, \bibinfo{address}{Berlin},
  \bibinfo{year}{2003}).

\bibitem[{\citenamefont{{T.\ Inagaki} et~al.}(1983)\citenamefont{{T.\ Inagaki},
  {M.\ Motosuga}, {K.\ Yamamori}, and {E.\ T.\ Arakawa}}}]{Inagaki}
\bibinfo{author}{\bibnamefont{{T.\ Inagaki}}},
  \bibinfo{author}{\bibnamefont{{M.\ Motosuga}}},
  \bibinfo{author}{\bibnamefont{{K.\ Yamamori}}}, \bibnamefont{and}
  \bibinfo{author}{\bibnamefont{{E.\ T.\ Arakawa}}}, \bibinfo{journal}{{Phys.\
  Rev.\ B }} \textbf{\bibinfo{volume}{28}}, \bibinfo{pages}{1740}
  (\bibinfo{year}{1983}).

\bibitem[{\citenamefont{{M.\ Iwanaga}}(2005)}]{Iwanaga}
\bibinfo{author}{\bibnamefont{{M.\ Iwanaga}}}, \bibinfo{journal}{{Phys.\ Rev.\
  B }} \textbf{\bibinfo{volume}{72}}, \bibinfo{pages}{012509}
  (\bibinfo{year}{2005}).

\bibitem[{\citenamefont{{K.\ G.\ Lee} and {Q-H.\ Park}}(2005)}]{Lee}
\bibinfo{author}{\bibnamefont{{K.\ G.\ Lee}}} \bibnamefont{and}
  \bibinfo{author}{\bibnamefont{{Q-H.\ Park}}}, \bibinfo{journal}{{Phys.\ Rev.\
  Lett.\ }} \textbf{\bibinfo{volume}{95}}, \bibinfo{pages}{103902}
  (\bibinfo{year}{2005}).

\end{thebibliography}

\end{document}